\documentclass[aps,prd,floats,floatfix,amssymb,twocolumn,superscriptaddress,nofootinbib]{revtex4-1}

\usepackage{graphicx}
\usepackage{ulem}
\usepackage{amssymb,amsmath}
\usepackage{graphicx}
\usepackage{color}
\usepackage{xfrac}
\usepackage{subcaption}
\usepackage{ulem}
\usepackage[dvipsnames]{xcolor}

\usepackage{lipsum}
\usepackage[font=footnotesize,labelfont={sf,bf}]{caption} 
\usepackage[font=footnotesize]{subcaption} 
\usepackage[export]{adjustbox} 
\usepackage{lipsum}
\usepackage{mwe}
\usepackage[margin=10pt,font=small,labelfont=bf,justification=raggedright]{caption}

\newcommand{\be}{\begin{equation}}
\newcommand{\ee}{\end{equation}}
\newcommand{\bse}{\begin{subequations}}
\newcommand{\ese}{\end{subequations}}
\newcommand{\eq}[1]{Eq.~(\ref{#1})}

\newcommand{\RR}{\mathbb{R}}

\begin{document}
\title{Quantum Mechanics on Sharply Bent  Wires via Two-Interval Sturm-Liouville Theory}

\author{Jo\~ao Paulo M. Pitelli}
\email[]{pitelli@unicamp.br}
\affiliation{Departamento de Matem\'atica Aplicada, Universidade Estadual de Campinas, 13083-859, Campinas, S\~ao Paulo, Brazil}

\author{Ricardo A. Mosna}
\email[]{mosna@unicamp.br}
\affiliation{Departamento de Matem\'atica Aplicada, Universidade Estadual de Campinas, 13083-859, Campinas, S\~ao Paulo, Brazil}

\author{Felipe Felix Souto}
\email[]{felipe\_felix\_souto@hotmail.com}
\affiliation{Departamento de Matem\'atica Aplicada, Universidade Estadual de Campinas, 13083-859, Campinas, S\~ao Paulo, Brazil}

\begin{abstract}

We study quantum mechanics on a curved wire by approximating the physics around the curved region by three parameters coming from the boundary conditions given by the two interval Sturm-Liouville theory.  Since the geometric potential on a  highly curved wire is strong an non-integrable,  these parameters depend on  the regularization of the curved wire.  Hence, unless we know precisely the shape of the wire, the presented method becomes not only a useful approximation, but also a necessary scheme to deal with quantum mechanics on highly curved wires.

\end{abstract}

\maketitle
\section{Introduction}
\label{introduction}

In contrast to what happens in classical mechanics, the quantization of systems with constraints  might lead to ambiguities if one is not careful in properly specifying the constraints defining the system.  In classical mechanics, the motion of a particle confined to a curve or surface  $M$ of the Euclidean space can be described in at least two different but equivalent ways.  The particle can be either confined to $M$ by transversal ever increasing constraining forces\footnote{There are some technicalities here though; one of them is that a tangent acceleration arises if the confining potential is not constant throughout the submanifold~\cite{van Kampen}).} or one might directly employ generalized coordinates avoiding the trouble of specifying the constraint forces.  In quantum mechanics, these two approaches lead to different results~\cite{Ikegami}, with the  first approach invariably introducing a geometric potential $V_g$ depending on the extrinsic curvature on $M$. This extra potential is completely missed by a direct naive application of the second approach.

As shown by da Costa in \cite{costa1,costa2}, given a smooth surface $S$ with principal curvatures $k_1$ and $k_2$,  the geometric potential is given by
\begin{equation}
V_g=-\frac{\hbar^2}{2m} (K-H^2),
\label{V_g formula}
\end{equation}
where $K=k_1k_2$ and $H=(k_1+k_2)/2$ are the Gaussian and mean curvatures of $S$, respectively.\footnote{We should emphasize that $V_g$ given by Eq.~(\ref{V_g formula}) is not a universal result. It depends crucially on the fact that the potential confining the particle is  uniform and transversal over the surface.} The effective Schr\"odinger equation which arises after eliminating the transversal mode of the wavefunction and taking the limit of infinite constraining force is then given by 
\begin{equation}
-\frac{\hbar^2}{2m}\Delta_{LB}\psi+V_g(q_1,q_2)\psi=i\hbar\frac{\partial \psi}{\partial t},
\end{equation}
where $q_1$, $q_2$ are coordinates on $S$, and $\Delta_{LB}$ is the corresponding covariant Laplacian (the Laplace-Beltrami operator). 
The case when the particle is confined to a wire, represented by a curve $C$ in $\RR^2$,  is completely analogous. Now, instead of $k_1$ and $k_2$, there is only one relevant curvature $k=1/R$, where $R$ is the radius of the osculating circle at a point of $C$. Moreover, if we parametrize $C$ by its arclength $s$, the covariant Laplacian reduces to $d^2/ds^2$ and the time-independent Schr\"odinger equation becomes
\[
-\frac{\hbar^2}{2m} \frac{d^2\psi}{ds^2}+V_g(s)\psi=E\psi,
\] 
with
\begin{equation}
\label{Vg}
V_g=-\frac{\hbar^2}{8m}k(s)^2=-\frac{\hbar^2}{8m}\frac{1}{R^2(s)}.
\end{equation}

\begin{figure}[h!]
\begin{center}
\includegraphics[width=0.25\textwidth]{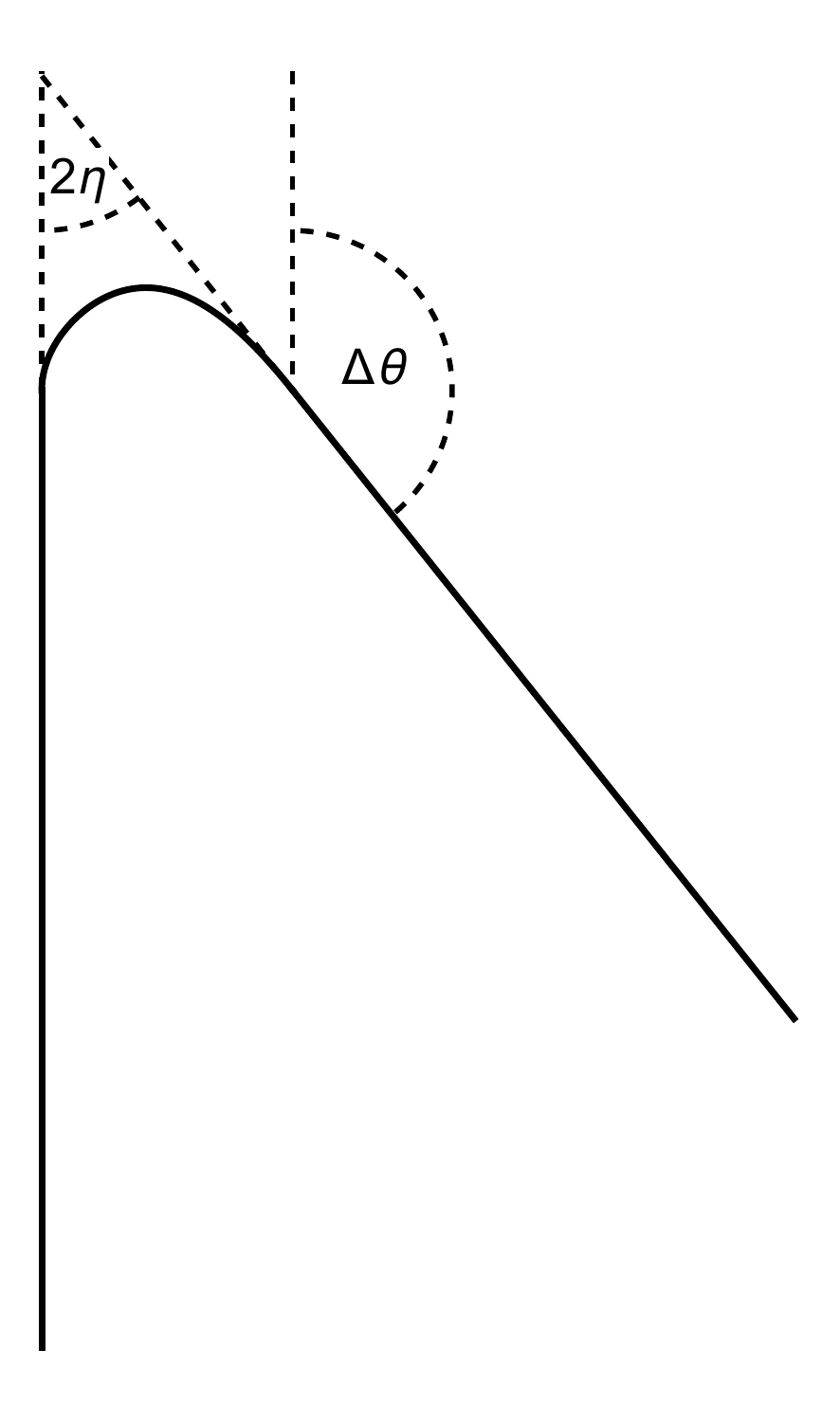}
\end{center}
\caption{A regularized corner (solid curve).}
\label{fig:bent_wire}
\end{figure}

This formulation works well if the constraints are smooth. However, the curvature is not well defined for corners (co-dimension $1$ sets on curves or surfaces) or tips (co-dimension $2$ sets on surfaces) and serious difficulties arise in the first case. This can be illustrated in the case of curves as follows. Consider a regularized version of the bent wire as in Fig.~\ref{fig:bent_wire}.  The only parameter describing the geometry of a sharply bent waveguide is its opening angle $2\eta=\pi-\Delta\theta$. Since the curvature $k$ of a curve gives the rate of angular increase of the curve's tangent vector, we have 
\be
\label{intcurv}
\int k(s) ds=\Delta\theta,
\ee 
where we integrate over the (small) region concentrating the curvature of the wire. In the limit of a sharp corner, we should thus have $k(s)=\Delta\theta\, \delta(s)$. On the other hand, we see from~\eq{Vg} that the geometric potential is proportional to $1/R^2(s)$ and therefore, in the limit of an abruptly bent curve, this behaves as a delta squared potential which is, of course, intrinsically ill defined.
This reflects the fact that, in the limit of a sharp corner, the geometric potential would be highly dependent on the way we regularize the curved region.  Different regularizations may lead to very different results.

\begin{figure}[h!]
\includegraphics[scale=0.55]{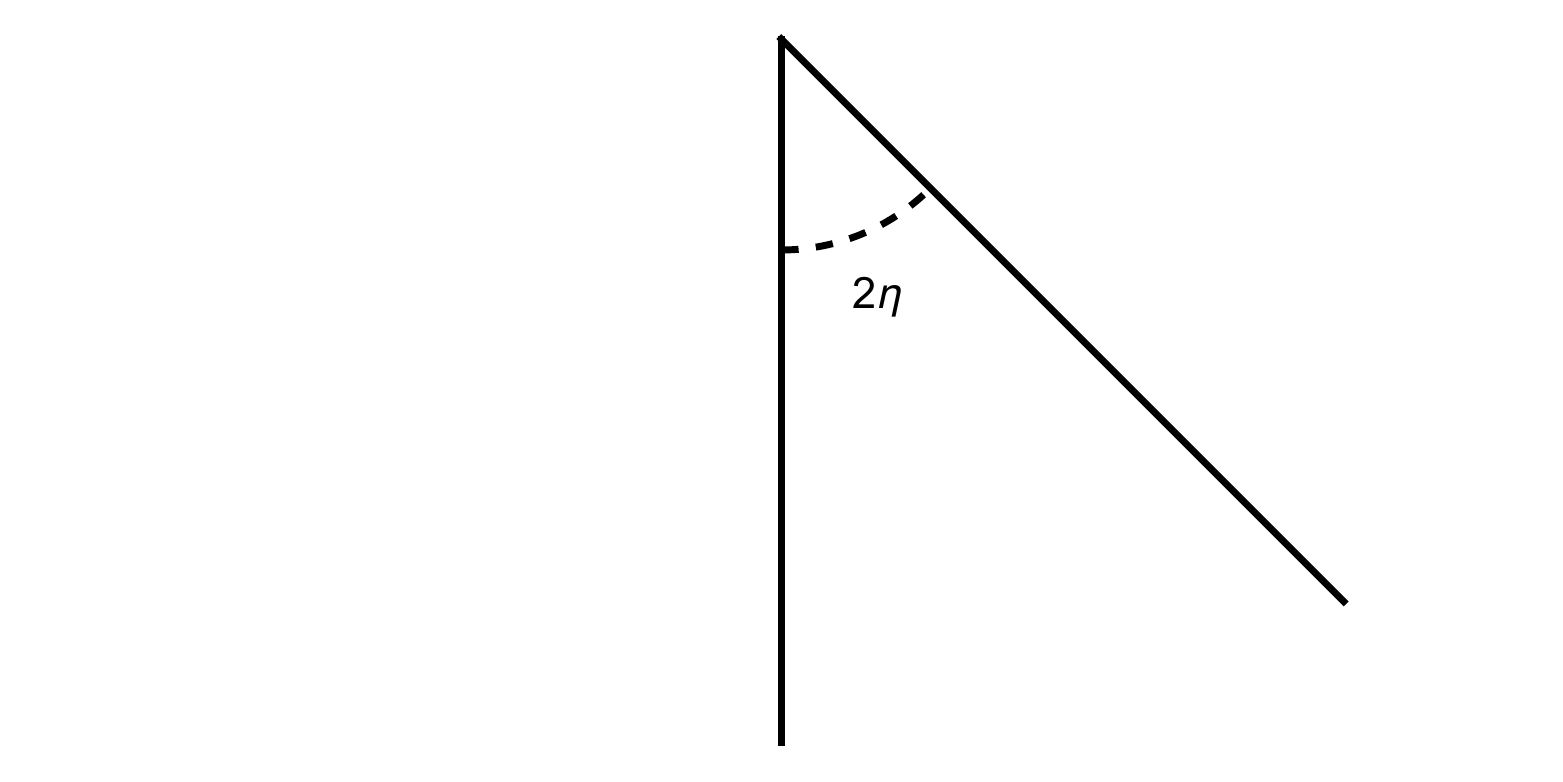}
\caption{Idealized bent wire.}
\label{wire idealized}
\end{figure}

Still, our intuition might tell us that it should be possible to extract some information for the physically relevant case of a sharply bent quantum wire.  We show that this is indeed the case when $kR\ll 1$,  where $E=\frac{\hbar^2k^2}{2m}$ is the energy of the quantum particle and $R$ is a length scale where  the curvature is significant.  We  will show that the quantum scattering on a sharply bent wire can be modelled by an idealized version of the corner as in Fig.~\ref{wire idealized} with appropriate boundary conditions.  Since quantum mechanics demands unitarity   (probability conservation), we will work only with self-adjoint boundary conditions.  These boundary conditions can be found via multi-interval Sturm-Liouville theory~\cite{zettl, cao}, which relates the values of the wave function (and its derivative) on both sides of the corner.

The use of self-adjoint boundary conditions to approximate lower dimensional quantum mechanics around highly curved regions was already  explored in Refs.~\cite{filgueiras,andrade}, where  quantum mechanics around the apex of a cone was considered. Contrarily to the case of the sharply bent wire, it can be shown by da Costa's formalism that the geometric potential in this case is a distribution and is given by
\begin{equation}
V_{\textrm{geo}}=-\frac{\hbar^2}{8m}\frac{1-\alpha^2}{\alpha^2 \rho^2}+\frac{\hbar^2}{2m}\frac{1-\alpha}{\alpha}\frac{\delta(\rho)}{\rho},
\end{equation}   
with $\rho$ being the radial coordinate and $\delta=2\pi(1-\alpha)$ the angular deficit.  Inspired by the seminal paper by Kay and Studer~\cite{kay}, the delta potential was treated as a short-range potential of the form $U_{\textrm{short}}\equiv\frac{\hbar^2}{2m}\frac{1-\alpha}{\alpha}\frac{\delta(\rho-a)}{a}$  with  $a\ll 1$.  Then,  the static  ($E=0$) Schr\"odinger equation was solved without this short-range potential with the help of the theory of self-adjoint extensions.  Finally, the arbitrary  self-adjoint extension parameter $\beta$ was related to the true problem by requiring that
\begin{equation}
\left.\frac{\rho}{\Psi^{\beta}_{\textrm{static}}}\frac{d\Psi^{\beta}_{\textrm{static}}}{d\rho}\right|_{\rho=a}=\left.\frac{\rho}{\Psi^{\textrm{true}}_{\textrm{static}}}\frac{d\Psi^{\textrm{true}}_{\textrm{static}}}{d\rho}\right|_{\rho=a},
\end{equation}
and noticing that the right-hand side of the above equation can be found by integrating the static   Schr\"odinger equation in the limit $a\to 0$. This procedure gives a unique relation between the unique self-adjoint extension parameter $\beta$ and the physics of the problem, represented by the parameter $\delta$. This was possible because the short-range potential could be integrated.  This relates to the fact that the physical parameters for a distributional potential do not depend on the particular regularization (see ~\cite{Gopalakrishnan}) for short length scales.

In this paper, we consider the much stronger (non distributional) geometric potential obtained by bending a wire into  a sharp corner.   In this case, the procedure of Kay and Studer~\cite{kay} cannot be used naively.  As a matter of fact,  since our potential is of the form  $\sim \delta^2(s)$,  we will find a set of (three) parameters characterizing each particular regularization of the wire.  This set of parameters (as opposed to the single self-adjoint parameter on the cone) comes from the fact that the ``singular'' submanifold  (the corner, in our case) has co-dimension $1$,  splitting the wire into two separate regions (this is the reason why we use  the two-interval Sturm-Liouville theory).

In summary, our main goal is the modelling of a sharply bent wire by an idealized corner with the help of the theory of self-adjoint  extensions.   We argue that,  even without absolute knowledge about  the regularization of the corner,  we can pretend we are working on an idealized corner with three experimental parameters.  We then find how these parameters depend on the opening angle and on the length scale $R$.  We will see that other physical observables, such as the bound state, can be predicted using these parameters as long as the energy scale $k$ satisfies $kR\ll 1$. We then conclude that it is possible to make physical predictions even when working with microscopic (or even nano-) wires, where control is lost on the exact way the wire curves.

This paper is organized as follows: in Sec.~\ref{two-interval}  we  briefly study the mathematical aspects of the two-interval Sturm-Liouville problem.  In Sec.~\ref{idealized corner}, we apply the formalism presented in Sec.~\ref{two-interval} to the case of an idealized corner and find how the transmission  and reflection amplitudes depend on the choice of boundary condition.  In Sec.~\ref{da costa} we study da Costa's formalism~\cite{costa1,costa2} for the case of  a curved wire.  We first argue that, for $kR\ll 1$, it is always possible to model the curved wire by an idealized corner using two-interval Sturm-Liouville theory. Then we illustrate our claim in two particular analytically solvable examples, namely the open book model and the exponential smoothed potential.  Finally,  we leave Sec.~\ref{conclusions} for our final considerations.

 \section{Two-interval Sturm-Liouville problem}
\label{two-interval}
 
Let  $-\infty\leq a_i<b_i\leq\infty$ and  $J_i=(a_i,b_i)$,  $i=1,2$,  with $b_1 \le a_2$.   Consider the Sturm-Liouville problem
\begin{equation}
-(p_iy'_i)'+q_i y_i=\lambda w_iy_i,\,\,\,\,i=1,2,
\label{sturm equation}
\end{equation}
where $w_i>0$ is the weight function  on each interval.  The appropriate Hilbert space  is given by $H=H_1+H_2$, with $H_i=L^2(J_i,w_i)$, $i=1,2$ and the  inner product between $f=\{f_1,f_2\}$ and $g=\{g_1,g_2\}$ in $H$ is defined by
\begin{equation}
(f,g)_H=(f_1,g_1)_{H_1}+(f_2,g_2)_{H_2}
\label{inner}
\end{equation}
with $f_i\in H_i$, $i=1,2$.   For the multi-interval Sturm-Liouville problem,  this is the correct inner product when probability amplitudes (and  norm of states),  as well as symmetry of operators defined in $H$,  are considered.

We say that an arbitrary point $c\in J_i\cup \{a_i,b_i\}$ is  in the limit circle case when both solutions of Eq.~(\ref{sturm equation}) are square-integrable around that point.  Otherwise, we say that $c$ is in the limit point case, where no boundary condition is necessary and square-integrability is sufficient to specify the solution around the point.  

In the case of two connected semi-infinite straight wires  as in Fig.~\ref{wire idealized},  we have $a_1=-\infty$, $b_1=0^{-}$,  $a_2=0^{+}$ and $b_2=\infty$. The parameters in the Sturm-Liouville equation~(\ref{sturm equation}) are given by $\lambda=\frac{2m E}{\hbar^2}$, $p_i=1$, $q_i=0$ and $w_i=1$,  $i=1,2$.  Therefore,  the time independent Schr\"odinger equation on each interval of the wire is given by
\begin{equation}
-\frac{d^2\Psi(s)}{ds^2}=\frac{2m E}{\hbar^2}\Psi(s),\,\,\,s\in(-\infty,0^{-})\cup(0^{+},\infty).
\label{sturm equation specific}
\end{equation}
The points  $a_1=-\infty$ and $b_2=\infty$ are in the limit point cases.  To deal with $a_2=b_1=0$,  we first note that these  are regular points of Eq.~(\ref{sturm equation specific}).  In this way, they are in the limit circle case.  Following the theory of two-interval Sturm-Liouville theory found in Ref.~\cite{zettl}, the boundary conditions respecting unitarity at these points are  given by
\begin{equation}
AY[a_2]+B Y[b_1]=0,
\label{eq bc1}
\end{equation}
with
\begin{equation}
Y[b_1]=\left[\begin{array}{c}
y_1(0^{-})\\
y_1'(0^{-})
\end{array}\right]\,\,\,\textrm{and}\,\,\,Y[a_2]=\left[\begin{array}{c}
y_2(0^{+})\\
y_2'(0^{+})
\end{array}\right], 
\end{equation}
and $A$ and $B$ being $2\times 2$ complex matrices satisfying
\begin{itemize}
\item[i)] $\textrm{rank}(A,B)=2$,
\item[ii)] $AEA^{\ast}-BEB^{\ast}=0$,
\end{itemize}
where $E=\left(\begin{array}{cc}
0 & -1\\
1 & 0
\end{array}\right)$.  It can be shown~\cite{zettl} that it is also possible to  represent Eq. ~(\ref{eq bc1}) with $A$ and $B$ respecting conditions i) and ii) in the canonical form
\begin{equation}
Y[a_2]=e^{i\gamma}K Y[b_1], \,\,\,-\pi<\gamma\leq\pi,\,\,K\in SL_2(\mathbb{R}),
\label{sa boundary condition}
\end{equation}
where $SL_2(\mathbb{R})$ is the group of the $2\times 2$ real matrices with  determinant one.
\section{Quantum Mechanics  on a Bent Wire}
\label{qm bent wire}
\subsection{Idealized Corner}
\label{idealized corner}
Let us consider the quantum scattering on an infinite wire  with length parameter $s\in(-\infty,\infty)$ sharply bent at $s=0$.  The  time-independent Schr\"odinger equation on the intervals $J_1=(-\infty,0)$ and $J_2=(0,\infty)$ is given by
\begin{equation}\begin{array}{cc}
-\frac{d^2}{ds^2}\psi_1(s)=k^2 \psi_1(s),&-\infty<s<0,\\
-\frac{d^2}{ds^2}\psi_2(s)=k^2 \psi_2(s),&0<s<\infty,
\end{array}
\end{equation}
with $k^2=\frac{2mE}{\hbar^2}$.

Let us consider a scattering problem with 
\begin{equation}\begin{array}{lc}
\psi_1(s)=e^{iks}+r e^{-iks},& -\infty<s<0,\\
\psi_2(s)=te^{iks},&0<s<\infty,
\end{array}
\end{equation}
where $t$ and $r$ are the transmission and reflection amplitudes, respectively.  By Eq.~(\ref{sa boundary condition}) we have
\begin{equation}
\left[\begin{array}{c}
\psi_1(0^{-})\\
\psi_1'(0^{-})
\end{array}\right]=e^{i\gamma}K\left[\begin{array}{c}
\psi_2(0^{+})\\
\psi_2'(0^{+})
\end{array}\right],
\label{sa bc}
\end{equation}
with $-\pi<\gamma<\pi$ and
\begin{equation}
K=\left(\begin{array}{cc}
a&b\\
c &\frac{1+bc}{a}
\end{array}\right)\in SL_2(\mathbb{R}),
\end{equation}
where we assumed $a\neq 0$, for simplicity.  A straightforward calculation shows that the reflection and transmission coefficients respecting the boundary conditions in Eq.~(\ref{sa bc}) are given by
\begin{equation}
\begin{array}{l}
r=-\frac{i a c+(a^2-bc-1) k+i a b k^2}{i a c+(a^2+bc+1) k-i a b k^2},\\
t=\frac{2 a e^{i \gamma } k}{i a c+(a^2+bc+1) k-i a b k^2}.
\end{array}
\label{transmission idealized}
\end{equation}
so that $|r|^2+|t|^2=1$. 

As an example, consider $\gamma=0$ and $K=\textrm{diag}(1,1)$. This choice is equivalent to the continuity of the wave function and its derivative, at $\sigma=0$ i.e., 
\begin{equation}
\psi_1(0^{-})=\psi_2(0^{+})\,\,\,\textrm{and}\,\,\,\psi'_1(0^{-})=\psi'_2(0^{+}).
\end{equation}
From Eq. ~(\ref{transmission idealized}), we see that $|r|^2=0$ and $|t|^2=1$  so that there is no reflected wave in this case.  

\begin{figure}[h!]
\includegraphics[scale=0.25]{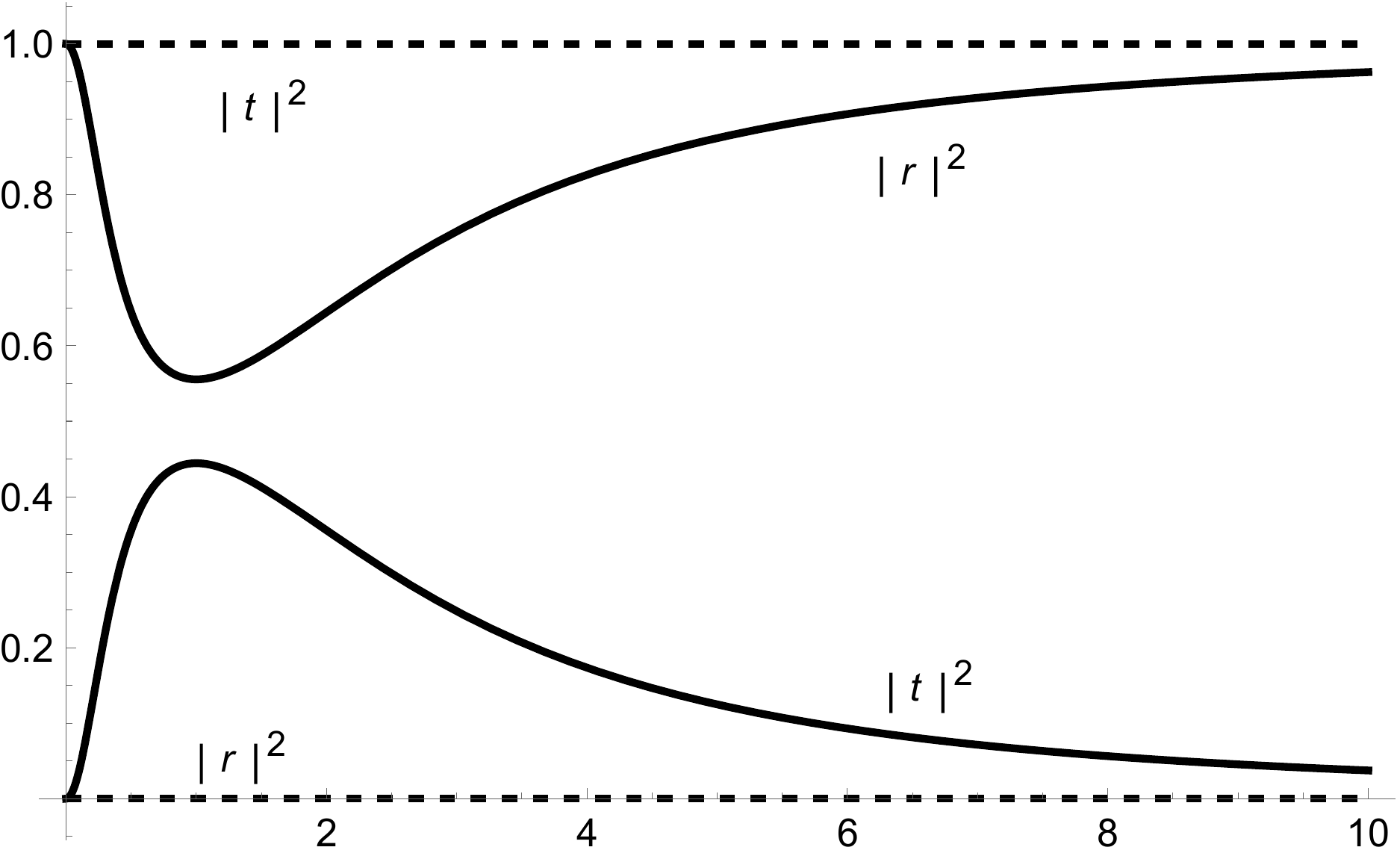}
\caption{Reflection and transmission amplitudes for  boundary conditions given by $\gamma=0$ and $K=\left(\begin{array}{cc}
1& 0\\
0 & 1
\end{array}\right)$ (dashed) and  $\gamma=0$ and $K=\left(\begin{array}{cc}
2& 1\\
1 & 1
\end{array}\right)$ (solid).}
\label{two examples}
\end{figure}

As another illustration, consider $\gamma=0$ and $K=\left(\begin{array}{cc}
2 & 1\\
1& 1
\end{array}\right)$, for instance.  We then have
\begin{equation}
|r|^2=1-\frac{4 k^2}{k^4+7 k^2+1},\,\,\textrm{and}\,\,\,|t|^2=\frac{4 k^2}{k^4+7 k^2+1}.
\end{equation}
We plot the transmission and reflection amplitudes as a function of $k$ for the two previous example in Fig.~\ref{two examples}. 

\subsection{Bent wire with geometric potential}
\label{da costa}

\subsubsection{Arbitrary shape}

Suppose that the particle is confined to a wire which is straight outside a small region of length $R$.  We also assume that this corner does not have any internal structure on a length scale below $R$. We see from \eq{Vg} that $k\sim 1/R$ so that $V_g\sim 1/R^2$.

Parametrizing $C$ by $s\mapsto \vec{r}(s)=\left(x(s),y(s)\right)$, where $s$ is the arc length of the curve, yields  
\[
V_g(s)=-\frac{\hbar^2}{8m}k^2(s)=-\frac{\hbar^2}{8m}\left\|\frac{d^2\vec{r}}{ds^2}\right\|^2,
\]
where we used the fact that $\|d\vec{r}(s)/ds\|=1$.  The scaling property of $V_g$ can be made manifest by expressing all the lengths in units of $R$. Writing $\sigma=s/R$ we have
\[
V_g=-\frac{\hbar^2}{8m}\frac{1}{R^4}\left\|\frac{d^2\vec{r}}{d\sigma^2}\right\|^2\equiv -\frac{\hbar^2}{8m}\frac{1}{R^2}U(\sigma),
\]
where $U(\sigma)\equiv \frac{\left\|\frac{d^2\vec{r}}{d\sigma^2}\right\|^2}{R^2}\leq 1$, since $R$ is our fundamental spatial scale.
In terms of the dimensionless coordinate $\sigma$, the Schr\"odinger equation becomes
\be
-\frac{d^2\phi}{d\sigma^2}+U(\sigma)\phi=R^2 k^2\phi,
\quad
\phi(\sigma)=R^{1/2} \psi(R \sigma)
\label{U de sigma}
\ee
(the prefactor ensures that $\phi$ is normalized like $\psi$; in particular, $\int|\phi(\sigma)|^2 d\sigma=1$ if $\int|\psi(s)|^2 ds=1$).

Consider two solutions $\phi_1(\sigma)$ and $\phi_2(\sigma)$ of Eq.~(\ref{U de sigma}) with energies $k_1$ and $k_2$, respectively. Since both $\phi_1$ and $\phi_2$ satisfy the Schr\"odinger equation for the same potential, we get
\[
\phi^{\ast''}_1  \phi_2-\phi_2''\phi_1^{\ast}=\left(k_2^2 R^2-k_1^2 R^2\right) \phi_1^{\ast}\phi_2.
\]
The left-hand side can be rewriten as $\frac{d}{d\sigma}\left(\phi_1^{\ast'}\phi_2-\phi_2'\phi_1^{\ast}\right)$ and we have, upon integration,
\begin{equation}
\begin{array}{l}
\left[\phi_1^{\ast'}(\sigma)\phi_2(\sigma)-\phi_2'(\sigma)\phi_1^{\ast}(\sigma)\right]_{-1/2}^{+1/2}=\\
R^2\left(k_2^2-k_1^2\right) \int_{-1/2}^{+1/2}\phi_1^{\ast}(\sigma)\phi_2(\sigma) d\sigma\\ \leq R^2\left(k_2^2-k_1^2\right),
\end{array}
\end{equation}
where, in the last line we used the Cauchy-Schwarz inequality. For $k_{i}$ satisfying $k_iR_i\ll 1$,  $i=1,2$,  we have 
\begin{equation}
\left[\phi_1^{\ast'}(\sigma)\phi_2(\sigma)-\phi_2'(\sigma)\phi_1^{\ast}(\sigma)\right]_{-1/2}^{+1/2}\approx 0.
\end{equation}
This equation holds  if
\begin{equation}
\begin{array}{l}
\phi_i(+1/2)=a \phi_i(-1/2)+b \phi_i'(-1/2),\\
\phi_i'(+1/2)=c \phi_i(-1/2)+d \phi_i'(-1/2)
\end{array},\,\,\,i=1,2,
\end{equation}
as long as
\begin{equation}
ad-bc=1.
\end{equation}
Note that $\sigma=\pm\frac{1}{2}$ implies that $s=\pm\frac{1}{2}R$. Also, $\psi_i$ is a function of the dimensionless parameter $ks$. In the limit $kR\ll 1$ we arrive at
\begin{equation}
\begin{array}{l}
\psi_i(0^{+})=a \psi_i(0^{-})+b \psi_i'(0^{-}),\\
\psi_i'(0^{+})=c \psi_i(0^{-})+d \psi_i'(0^{-})
\end{array},\,\,\,i=1,2,
\end{equation}
which is the condition given by Eq.~(\ref{sa boundary condition}) with $\gamma=0$.  We then conclude that, in this case,  regardless  of the exact form of the potential on the curved region, we can model the wire as a perfect corner as long as $kR\ll 1$. 

Let us illustrate our last statement with two exact solvable potentials.

\subsubsection{Open book Shape}

\begin{figure}[h]
\begin{center}
\includegraphics[width=0.25\textwidth]{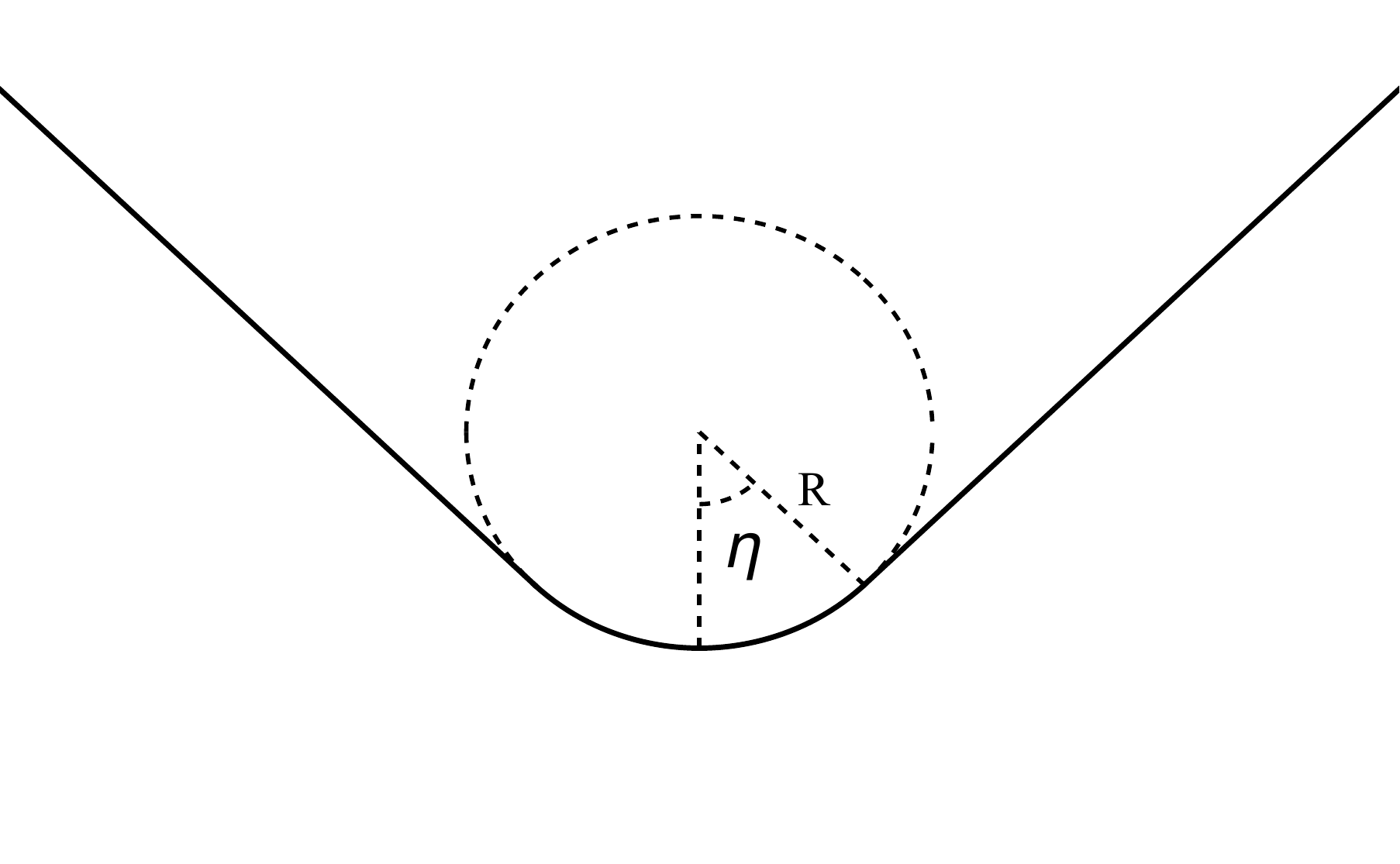}
\end{center}
\caption{Ilustration of the open book  model.}
\label{fig:wiggles}
\end{figure}

Consider the regularization shown in Fig. \ref{fig:wiggles}. The non-straight portion of the wire is limited to a (small) circular region of length $2R\eta$.   It follows from~\eq{Vg} that
\[
 \frac{\hbar^2}{2m}   V_g(s)= 
\begin{cases}
    0,& \text{if } |s|>R\eta\\
    -\frac{1}{4}\frac{1}{R^2},& \text{if } |s|<R\eta,\\
\end{cases}
\]

The time-independent Schr\"odinger equation becomes
\begin{equation}\begin{array}{lc}
-\frac{d^2}{ds^2}\phi_1(s)=k^2\phi_1(s),&-\infty<s<-R\eta,\\
-\frac{d^2}{ds^2}\phi_2(s)-\frac{1}{4R^2}\phi_2(s)=k^2\phi_2(s),&-R\eta<s<R\eta,\\-\frac{d^2}{d\sigma^2}\phi_3(s)=k^2\phi_3(s),&R\eta<s<\infty.
\end{array}
\end{equation}

Let us consider a scattering problem such that
\begin{equation}
\begin{array}{lc}
\phi_1(s)=e^{i k s}+re^{-i k s},&-\infty<s<R\eta,\\
\phi_2(s)=\alpha e^{i s\frac{ \sqrt{1+4 R^2 k^2}}{2R}}+\beta e^{-i s\frac{ \sqrt{1+4 R^2 k^2}}{2 R}},&-R\eta<s<R\eta,\\
\phi_3(s)=t e^{i k s},&R\eta<s<\infty.
\end{array}
\end{equation}
By demanding continuity of the wave function and its derivative at $s=\pm R\eta$, we arrive at\begin{widetext}
\begin{equation}
\begin{array}{l}
r=\frac{e^{-2 i \eta  k R} \left(-1+e^{2 i \eta  \sqrt{4 k^2 R^2+1}}\right)}{-8 k^2 R^2 e^{2 i \eta  \sqrt{4 k^2 R^2+1}}+4 k R \sqrt{4 k^2 R^2+1} e^{2 i \eta  \sqrt{4 k^2 R^2+1}}-e^{2 i \eta  \sqrt{4 k^2 R^2+1}}+8 k^2 R^2+4 k R \sqrt{4 k^2 R^2+1}+1},\\
t=\frac{8 k R \sqrt{4 k^2 R^2+1} e^{i \eta  \sqrt{4 k^2 R^2+1}-2 i \eta  k R}}{-8 k^2 R^2 e^{2 i \eta  \sqrt{4 k^2 R^2+1}}+4 k R \sqrt{4 k^2 R^2+1} e^{2 i \eta  \sqrt{4 k^2 R^2+1}}-e^{2 i \eta  \sqrt{4 k^2 R^2+1}}+8 k^2 R^2+4 k R \sqrt{4 k^2 R^2+1}+1}.
\end{array}
\label{transmission da costa open book}
\end{equation}
\end{widetext}
Notice that these coefficients depend on the specific regularization of the bent wire (in this case an open book model with parameters $R$ and $\eta$). However, let us consider the low energy scattering  $Rk\ll 1$ and expand $t$ and $r$ up to $\mathcal{O}(k^2R^2)$.  We have
\begin{widetext}
\begin{equation}
\begin{array}{l}
r=-1+2 i k R \left(\eta +2 \cot{\eta}\right)+2 k^2R^2  \left[\eta ^2+4 \eta  \cot{\eta}+2 i k R(\eta +2 \cot{\eta})+8 \csc ^2{\eta}-4\right],\\
t=4 i k  R \csc{\eta}+8k^2 R^2  \left(\eta +2 \cot{\eta}\right)) \csc{\eta}.
\end{array}
\label{da costa series}
\end{equation}
\end{widetext}

By expanding Eq.~(\ref{transmission idealized}) up so second order in $k$ and comparing term by term with Eq.~(\ref{da costa series}), we find
\begin{equation}
\begin{array}{ll}
a=\frac{1}{2} \eta  \sin{\eta}+\cos{\eta},\\
b=-\frac{1}{2} R \sin{\eta} \left(\eta ^2+4 \eta  \cot{\eta}-4\right),\\
c=-\frac{\sin{\eta }}{2R}.
\end{array}
\label{copy}
\end{equation}

\begin{figure}[h!]
\includegraphics[scale=0.2]{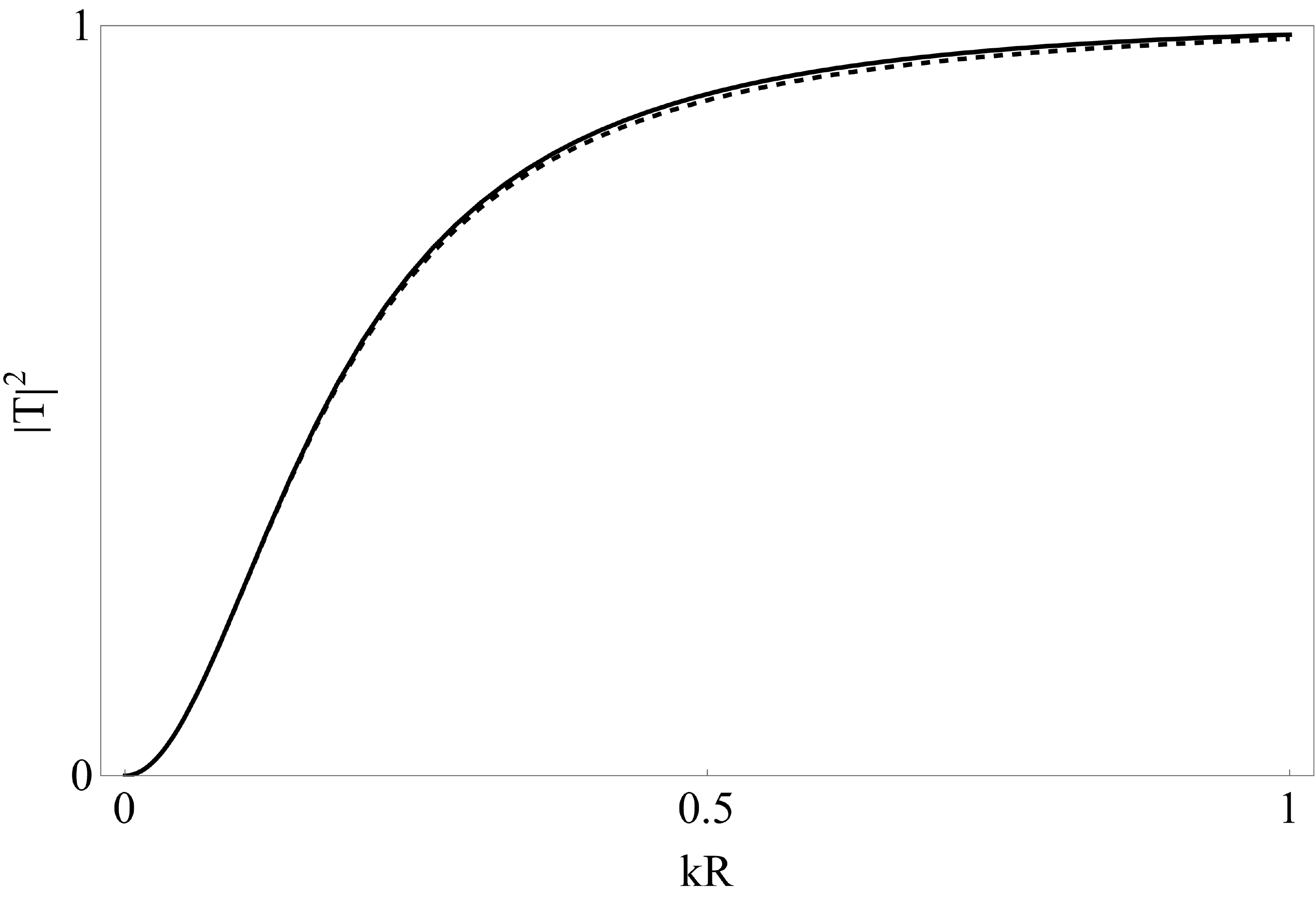}
\caption{Transmission coefficient as a function of $kR$ for the open book shaped wire (dashed) and for the idealized bent wire with $a$,  $b$ and $c$ given by Eq.~(\ref{copy}) (olid). We chose $\eta=\pi/4$.}
\label{fig wire idealized da costa}
\end{figure}

In Fig.~\ref{fig wire idealized da costa} we plot the transmission coefficient $|t|^2$ as a function of $kR$ for the open book model and for the idealized bent wire with boundary condition parameters satisfying Eq.~(\ref{copy}).  We see that our modelling works well for $kR\lesssim 1$. 

\begin{figure}[h!]
\includegraphics[scale=0.2]{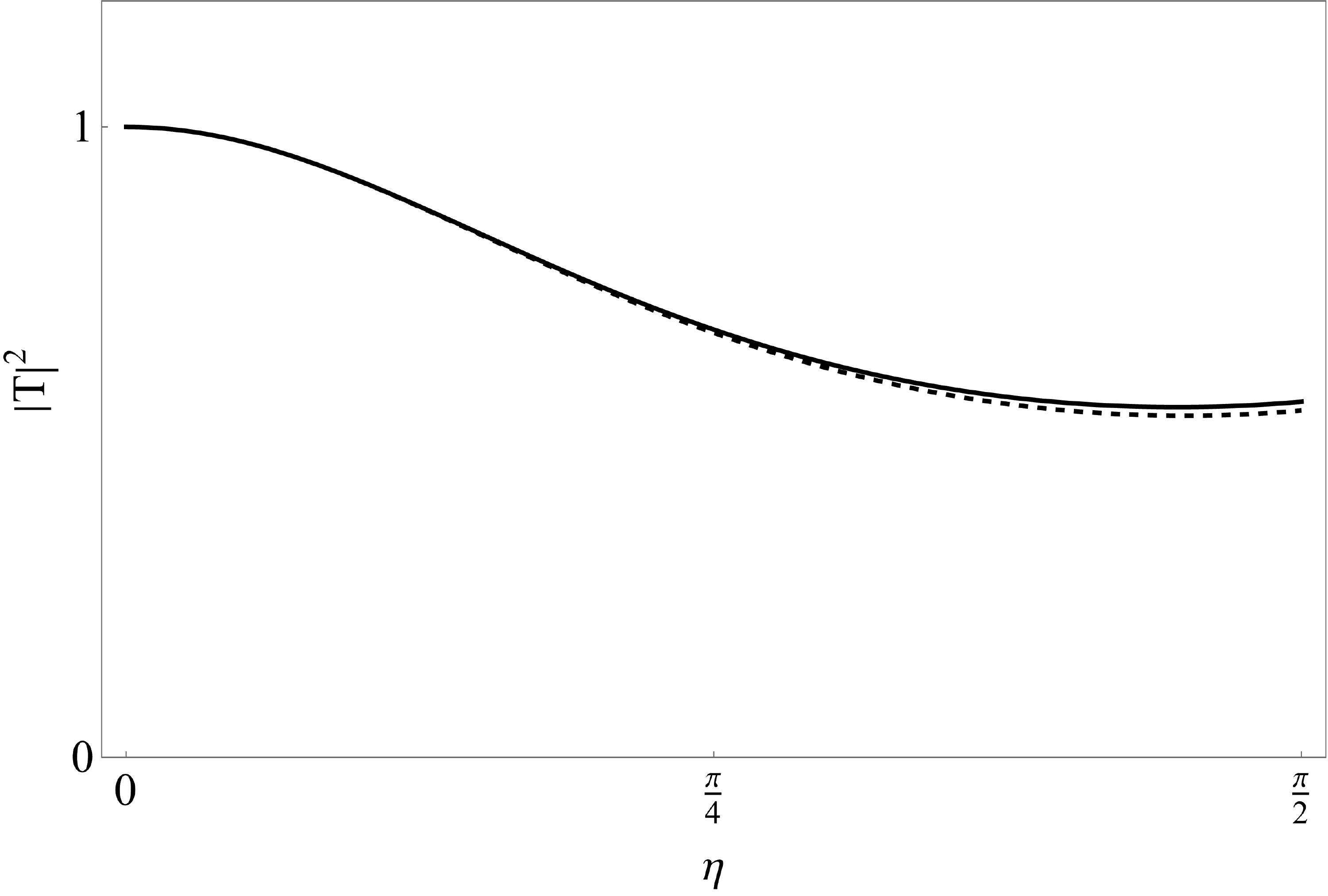}
\caption{Transmission amplitude as a function of $\eta$ for the open book shaped wire (solid) and for the idealized bent wire with $a$,  $b$ and $c$ given by Eq.~(\ref{copy}) (dashed). We chose $kR=1/4$.}
\label{fig wire idealized da costa angle}
\end{figure}

The other crucial parameter is the angle $\eta$. In Fig.~\ref{fig wire idealized da costa angle} we illustrate how our modelling depends on $\eta$ for $kR$ fixed.
In summary, Figs.~\ref{fig wire idealized da costa} and \ref{fig wire idealized da costa angle} show that for particles with small enough energies (with respect to $1/R^2$) and small deflection angles $\eta$, our idealized model works well.

\begin{figure}[h!]
\includegraphics[scale=0.20]{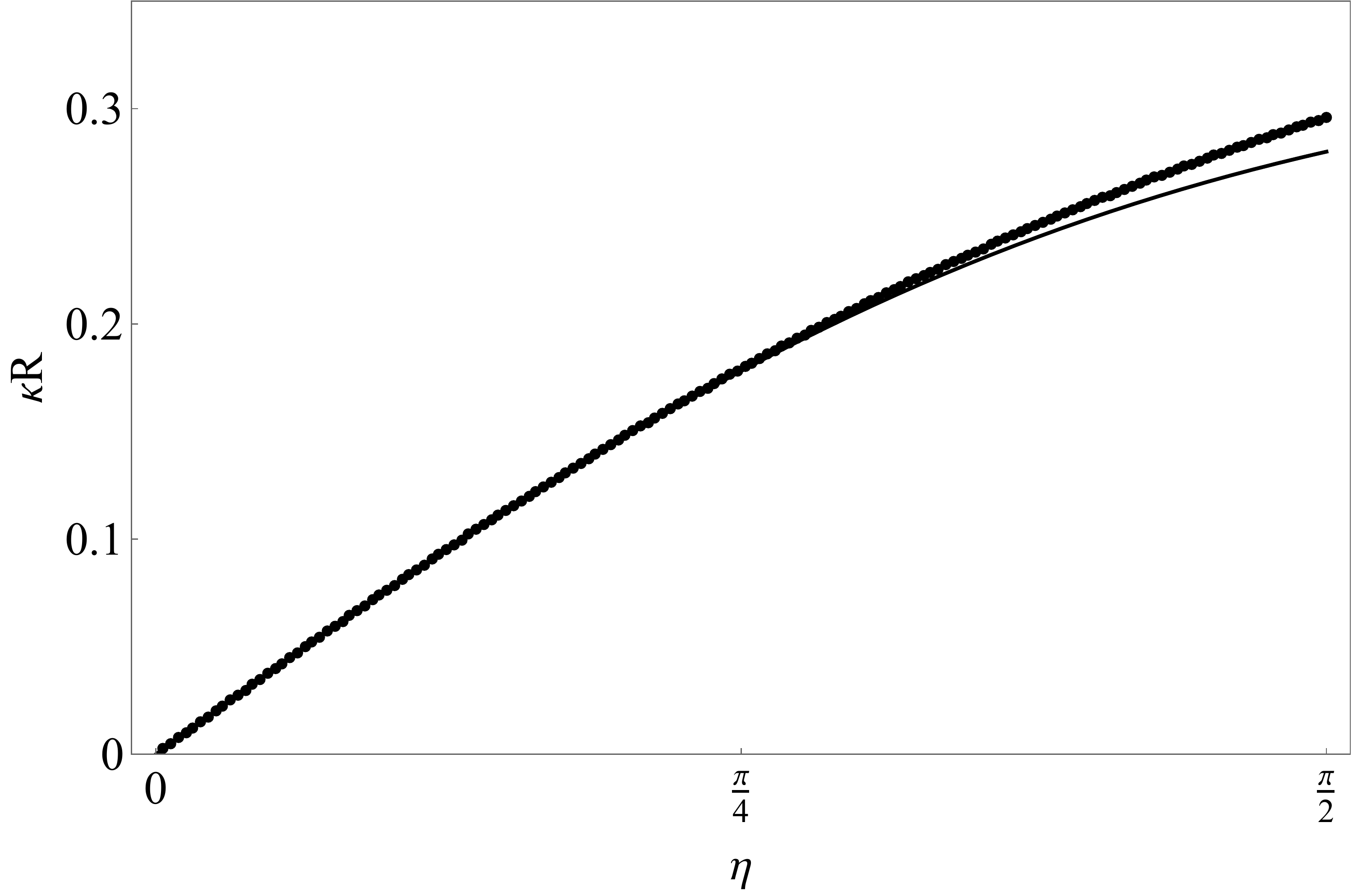}
\caption{Bound state for the open book (solid)  and the idealized models (dashed). Once more, we notice that our model works well for $\kappa R\ll 1$ (small angles). }
\label{bound open book}
\end{figure}

Given the coefficients in Eq.~(\ref{copy}) obtained by a scattering experiment, we wonder whether other physical observables can be predicted using this information. Let us look, for instance, at the bound state for the open book model. It is given by the value $\kappa$ for which
\begin{equation}
\begin{array}{lc}
\phi_1(s)=e^{\kappa s},&-\infty<s<R\eta,\\
\phi_2(s)=\alpha e^{s\frac{ \sqrt{-1+4 R^2 \kappa^2}}{2R}}+\beta e^{-s\frac{ \sqrt{-1+4 R^2 \kappa^2}}{2 R}},&-R\eta<s<R\eta,\\
\phi_3(s)=e^{-\kappa s},&R\eta<s<\infty.
\end{array}
\end{equation}
with $\alpha$ and $\beta$ extracted by the continuity of $\phi$ and $\phi'$ at $s=\pm R\eta$.  A straightforward  calculation shows that 
\begin{widetext}
\begin{equation}\begin{aligned}
\frac{1}{\kappa R \sqrt{4 \kappa^2R^2-1}}&\bigg[e^{-\left(\eta  \left(\sqrt{4 R^2\kappa^2-1}+\kappa R\right)\right)} \left(\left(2 \kappa R \left(\sqrt{4 \kappa^2R^2-1}-2 \kappa R\right)+1\right) e^{\eta  \kappa R}\right.\\&+\left.\left(4 \kappa R \left(\sqrt{4 \kappa^2R^2-1}+2 \kappa R\right)-1\right) e^{\eta  \left(2 \sqrt{4 \kappa^2R^2-1}+\kappa R\right)}+2 \kappa R \left(\sqrt{4 \kappa^2R^2-1}-2 \kappa R\right)\bigg]\right.=0.
\end{aligned}
\end{equation}
\end{widetext}
We solved the above equation numerically using the software Mathematica~\cite{mathematica}. We plot $\kappa R$ as a function of the angle $\eta$ in Fig.~\ref{bound open book}. There, we also plot $\kappa R$ for the bound state of the idealized model with coefficients given by Eq.~(\ref{copy}). A simple calculation shows that it is given by
\begin{equation}\begin{aligned}
\kappa R&=-\frac{-\sqrt{\left(-\frac{b c}{a}-a-\frac{1}{a}\right)^2-4 bc}+\frac{bc}{a}+a+\frac{1}{a}}{2 b}\\&=\frac{1}{\eta + 2 \cot{(\eta/2)}}
\end{aligned}\end{equation}
We see that, for small opening angle $\eta$ (or, equivalently, $\kappa R\ll 1$), our model predicts the bound state with a very high degree of accuracy.

\subsection{Exponentially Smoothed Potential}

\begin{figure}
\includegraphics[scale=0.25]{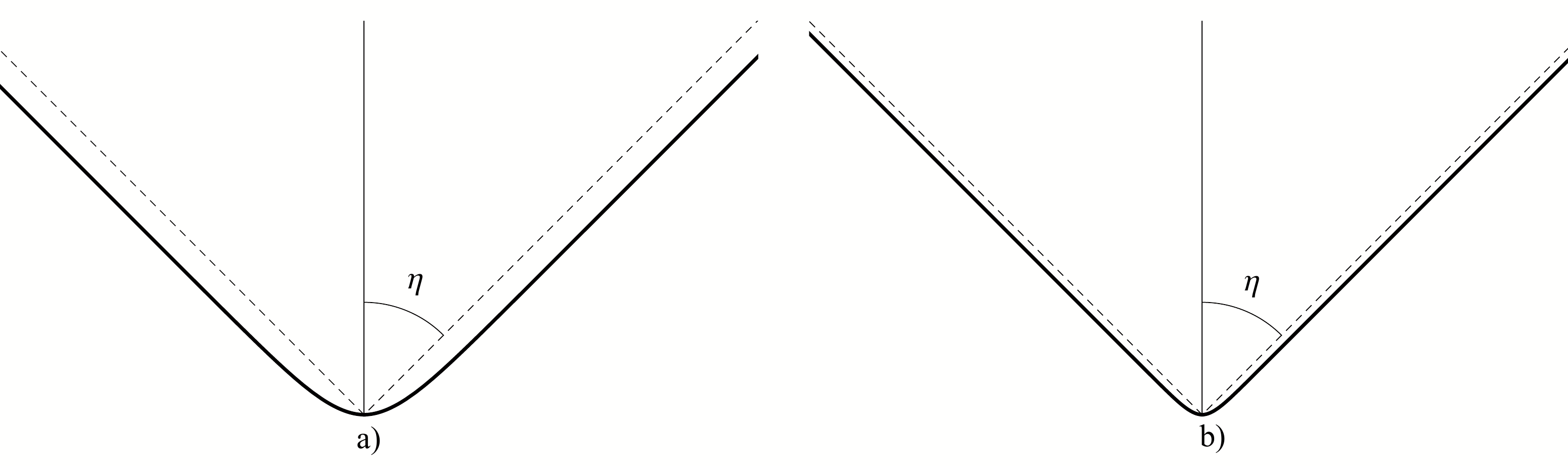}
\caption{The form of the regularized wire for a) $\Lambda=0.3$ and b) $\Lambda=0.1$. In both cases $\eta=\pi/4$.}
\label{figexp}
\end{figure}

We can also work with a different exact solvable smoothing potential. A potential of the form $U(s)=-\lambda e^{-|s|/\Lambda}$, where $s$ is the arc length for the regularized curve modeling the V-shaped wire, fills this purpose \cite{fabre}. In order to model the wedge, the coefficient $\lambda$ must be of the form $\lambda=\eta^2/(16\Lambda^2)$, where $\eta$ is the angle of the wedge and $\Lambda$ is the regularizing parameter, as can be seen in  Fig. \ref{figexp}.

Starting with an incoming plane wave coming from $s\to-\infty$ we have the scattering solution
\begin{equation}
\psi_{s<0}=\frac {J_ {2 i \Lambda k}\left (\frac{1}{2} e^{\frac{s}{2  \Lambda}}\eta \right)} {\eta^{2 i  \Lambda k}}+r\frac { J_ {-2 i  \Lambda k}\left (\frac {1} {2} e^{\frac {s} {2  \Lambda}}\eta \right)} {\eta^{-2 i  \Lambda k}},
\end{equation}
for $s<0$ and 
\begin{equation}
\psi_{s>0}=t\, \eta^{2 i  \Lambda k} J_ {-2 i  \Lambda k}\left (\frac {1} {2} e^{\frac {s} {2 \Lambda}}\eta \right),
\end{equation}
for $s>0$. In the above equations, $t$ and $r$ are the transmission and reflection amplitudes,  respectively.

Matching the usual conditions of continuity of the wave function and its derivative we arrive at\begin{widetext}
\begin{equation}\begin{array}{l}
r=\frac{2^{-1+8 i k  \Lambda } \eta ^{-4 i k  \Lambda } \Gamma(2 i k \Lambda +1) \left(\frac{J_{2 i k  \Lambda +1}\left(\frac{\eta }{2}\right)-J_{2 i k  \Lambda -1}\left(\frac{\eta }{2}\right)}{J_{-2 i k  \Lambda -1}\left(\frac{\eta }{2}\right)-J_{1-2 i k  \Lambda}\left(\frac{\eta }{2}\right)}-\frac{J_{2 i k  \Lambda }\left(\frac{\eta }{2}\right)}{J_{-2 i k  \Lambda}\left(\frac{\eta}{2}\right)}\right)}{\Gamma (1-2 i k  \Lambda )},\\
t=-\frac{2^{1+8 i k  \Lambda } \eta ^{-4 i k  \Lambda } \sinh (2 \pi  k  \Lambda) \Gamma(2 i k  \Lambda)}{\pi  \Gamma(-2 i k  \Lambda ) J_{-2 i k \Lambda}\left(\frac{\eta }{2}\right) \left(4 k  \Lambda  J_{-2 i k  \Lambda }\left(\frac{\eta }{2}\right)-i \eta  J_{1-2 i k  \Lambda }\left(\frac{\eta }{2}\right)\right)}.
\end{array}
\end{equation}
\end{widetext}

\begin{figure}[h!]
\includegraphics[scale=0.25]{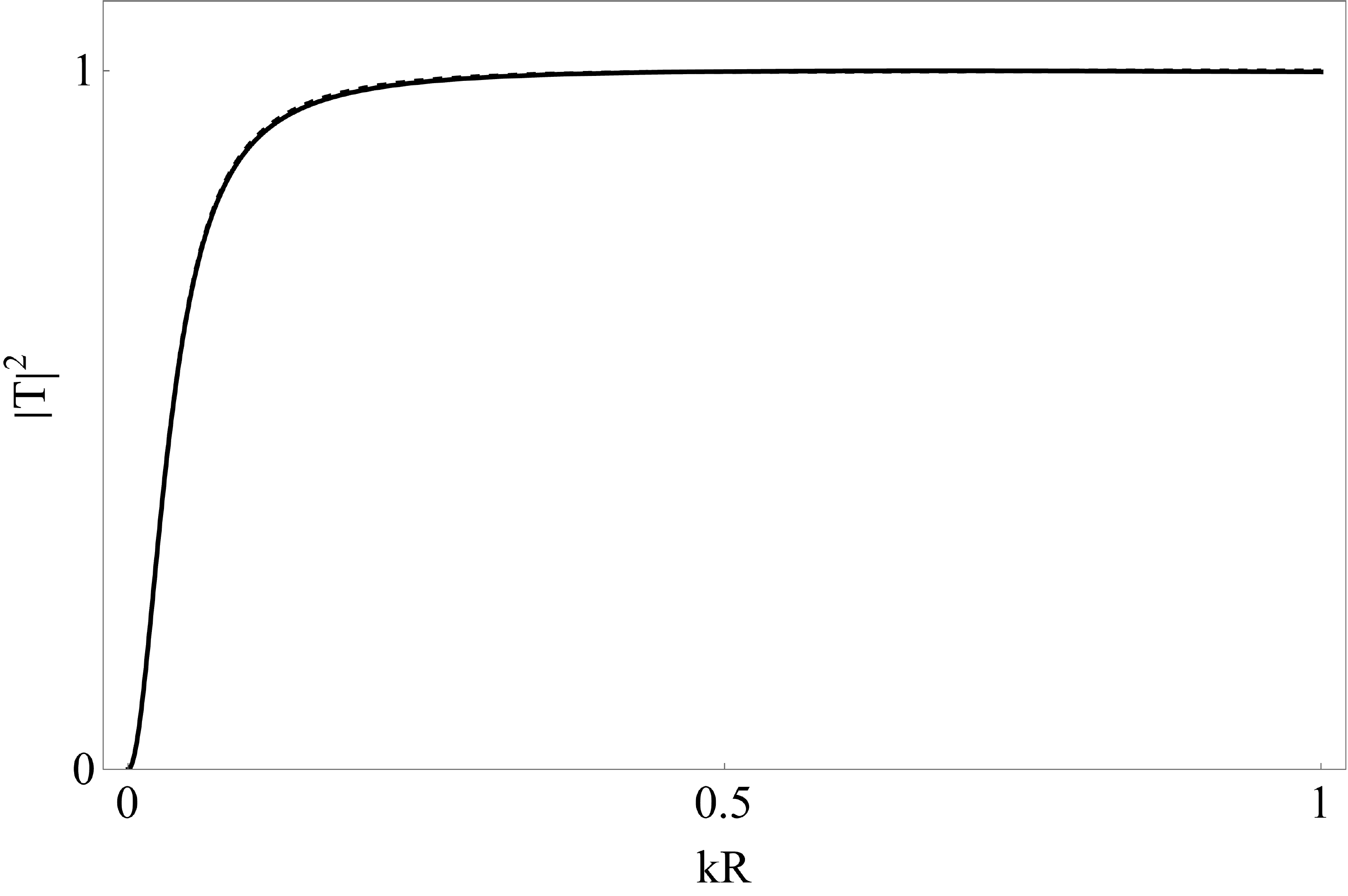}
\caption{Transmission coefficient as a function of $k R$  for the exponential smoothed potential (solid) and for the idealized bent wire with $a$, $b$ and $c$ given by Eq.~(\ref{coefficients exponential}) (dashed) . We chose $\eta=\pi/4$.}
\label{transmission exponential}
\end{figure}

\begin{figure}[h!]
\includegraphics[scale=0.25]{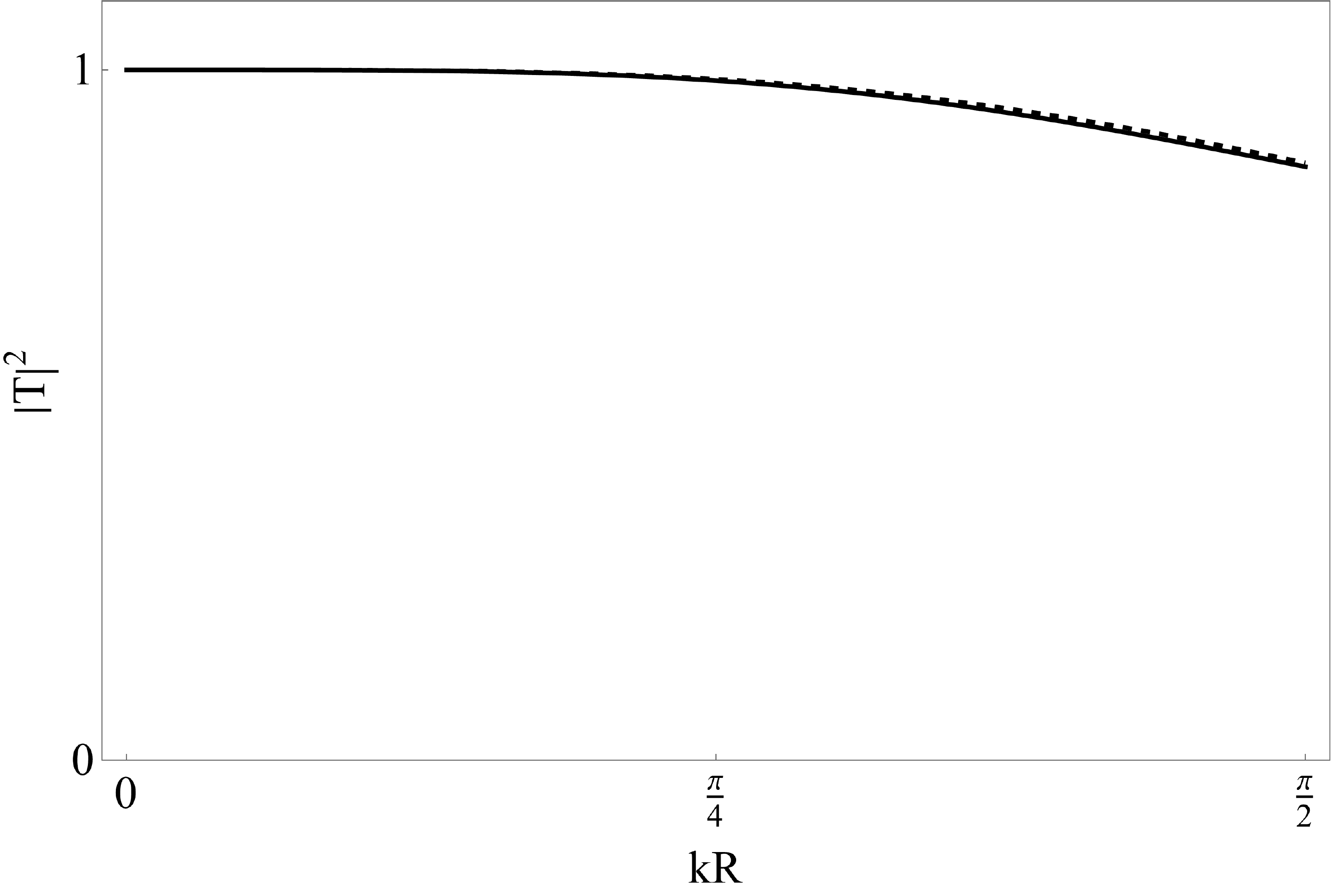}
\caption{ransmission amplitude as a function of $\eta$ for the exponential smoothed potential (solid) and for the idealized bent wire with $a$,  $b$ and $c$ given by Eq.~(\ref{copy}) (dashed). We chose $kR=1/4$.}
\label{transmission exponential angle}
\end{figure}

\begin{figure}[h!]
\includegraphics[scale=0.25]{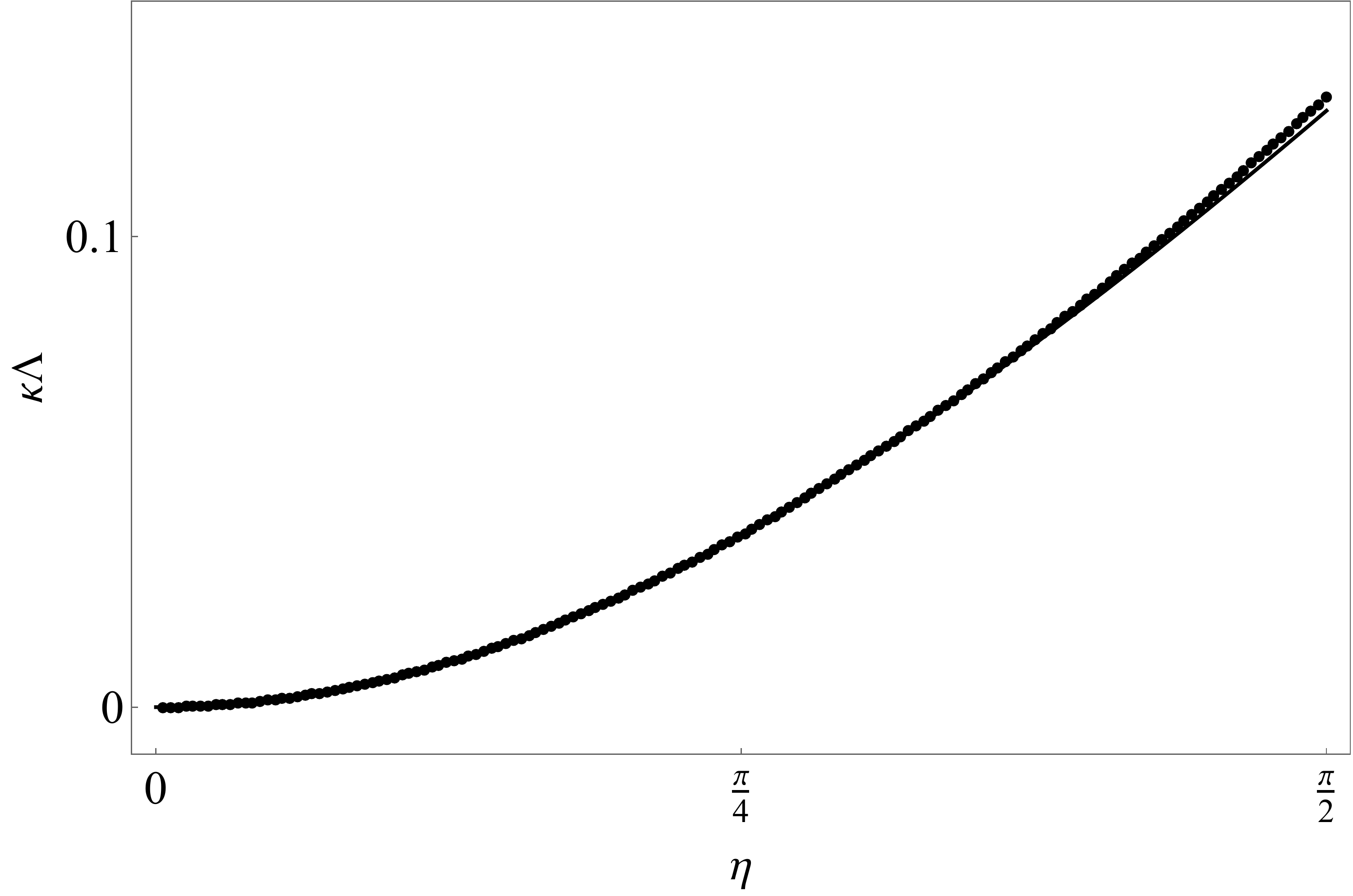}
\caption{Bound state for the  exponential smoothed potential (solid) and the idealized model. (dashed) Once more, we notice that our model works well for $\kappa R\ll 1$ (small angles). }
\label{bound exponential}
\end{figure}

By the same procedure employed in the case of the open book model, we find\begin{widetext}
\begin{equation}
\begin{array}{l}
a=\frac{1}{4} \eta  \left(J_0\left(\frac{\eta }{2}\right) \left(4 \left(\log \left(\frac{\eta }{4}\right)+\gamma \right) J_1\left(\frac{\eta }{2}\right)-\pi  Y_1\left(\frac{\eta }{2}\right)\right)-\pi  J_1\left(\frac{\eta }{2}\right) Y_0\left(\frac{\eta }{2}\right)\right),\\
b=\frac{1}{2} \eta  \sigma  \left(2 \left(\log \left(\frac{\eta }{4}\right)+\gamma \right) J_0\left(\frac{\eta }{2}\right)-\pi  Y_0\left(\frac{\eta }{2}\right)\right) \left(\pi  Y_1\left(\frac{\eta }{2}\right)-2 \left(\log \left(\frac{\eta }{4}\right)+\gamma \right) J_1\left(\frac{\eta }{2}\right)\right),\\
c=-\frac{\eta  J_0\left(\frac{\eta }{2}\right) J_1\left(\frac{\eta }{2}\right)}{2  \Lambda }.
\end{array}
\label{coefficients exponential}
\end{equation}
\end{widetext}
In Fig.~(\ref{transmission exponential}) we show the transmission coefficient as a function of $k R$ for the smoothed exponential case and the idealized model with coefficients given by Eq.~(\ref{coefficients exponential}).
The transmission coefficient (for fixed $kR$) and the bound state as a function of the angle $\eta$ are shown in Figs.~\ref{transmission exponential angle} and \ref{bound exponential}.


\section{Conclusions}
\label{conclusions}
Quantum mechanics constrained to a curved wire with curvature scale $\sim1/R$ can be approximated by an idealized bent wire with appropriate boundary conditions. The approximation works  as long along  as $kR\ll 1$, where $k$ is the energy scale.  Since the geometric potential  coming from da Costa's formalism is of the form $\sim \delta^2(s)$ in the limit $R\to 0$,  the boundary conditions parameters will depend not only on the curvature scale and on the the opening angle, but also on the particular way the wire curves (i.e., on the particular regularization of the sharply bent region).  In this way, for micro (or even nano-) wires, where the exact curvature is unknown  we must use the approximation presented above to have an appropriate description for the region $kR\ll 1$.

The approach presented here is a new way of describing quantum mechanics on manifolds with singularities of co-dimension 1.  In Refs.~\cite{filgueiras,andrade}, the theory of self-adjoint extensions was also used to describe quantum mechanics around the apex of a cone.  However, since the geometric potential was integrable and the singular region was a single point on the conical surface, there was a single parameter related to the boundary condition which could be related to the physical parameters in a single way.  On the other hand,  the potential generated by the curvature singularity on a sharply bent wire is non-integrable and splits the wire into two disjoint intervals.  As a result, three independent boundary condition parameters become necessary to describe the curved region.  Furthermore,  these parameters cannot be related to the physical parameter (namely,  the angle) in a single way. They depend crucially on the regularization scheme. This can be seen by comparing Figs.~\ref{fig wire idealized da costa} and~\ref{transmission exponential}. The transmission amplitude are completely different for the open book and exponentially smoothed potential, even though the two regularizations lead to the same limiting curved wire. In this way, our method becomes not only a mere approximation, but an effective way of deal with sharp curved wires.   

\acknowledgements
J.P.M.P. was partially supported by Conselho Nacional de
Desenvolvimento Científico e Tecnológico (CNPq, Brazil) under Grant No.
311443/2021-4. R. A. M. was partially supported by Conselho Nacional de Desenvolvimento Científico e Tecnológico (CNPq, Brazil) under Grant No. 310403/2019-7.  F.  F.  S. was supported by Coordenação de Aperfeiçoamento
de Pessoal de Nível Superior (CAPES, Brazil) under Grant No.  001-1684974.

\end{document}